\documentclass{JHEP3} 

\usepackage{multicol,bbm}
\usepackage{graphicx}
\usepackage{subfigure}
\newcommand\fverb{\setbox\fverbbox=\hbox\bgroup\verb}
\newcommand\fverbdo{\egroup\medskip\noindent%
            \fbox{\unhbox\fverbbox}\ }
\newcommand\fverbit{\egroup\item[\fbox{\unhbox\fverbbox}]}
\newbox\fverbbox

\newcommand{\be}{\begin{equation}}
\newcommand{\ee}{\end{equation}}
\newcommand{\bea}{\begin{eqnarray}}
\newcommand{\eea}{\end{eqnarray}}
\newcommand{\ba}{\begin{array}}
\newcommand{\ea}{\end{array}}

\def\a{\alpha}
\def\b{\beta}

\def\o{\omega}

\def\r{\rho}

\def\t{\tau}

\def\O{\Omega}

\title{Nuclear matter to strange matter transition in holographic QCD}

\author{Youngman Kim
        \\
Asia Pacific Center
for Theoretical Physics and Department of Physics, Pohang University
of Science and Technology, Pohang, Gyeongbuk 790-784, Korea
\\  E-mail: \email{ykim@apctp.org}}
\author{Yunseok Seo
\\
Center for Quantum Spacetime, Sogang University, Seoul 121-742, Korea
\\ E-mail: \email{yseo@sogang.ac.kr}}
\author{Sang-Jin Sin
\\
Department of Physics Hanyang University, Seoul 133-791, Korea
\\ E-mail: \email{sjsin@hanyang.ac.kr}}
\received{\today}       

\abstract{
We construct a simple holographic QCD model to study nuclear matter to strange matter transition.
The interaction of dense medium and hadrons is taken care of by
imposing the force balancing condition for stable D4/D6/D6 configuration.
By considering the intermediate and light flavor branes interacting with baryon vertex
homogeneously distributed along $R^3$ space and requesting the energy minimization, we
find that {\it there is a  well defined transition density}  as a function of  current quark mass.
We also find that as density goes up very high, intermediate (or heavy) and light quarks
populate equally as expected from the Pauli principle.
In this sense, the effect of the Pauli principle is realized as dynamics of D-branes.}


\begin{document}

\section{Introduction}

One of the challenging current problems in hadron physics is to elucidate the behavior of dense
matter under extremely high-density environments, for a review, e.g. see~\cite{denseM}.
It is theoretically
expected that, at very high baryon densities (even at low temperatures), chiral symmetry is
likely to be restored, and that baryon matter can be converted into quark matter.
Various studies also suggest the possible formation of a kaon condensate
at high densities. The existence of quark matter and/or a kaon condensate
can have important consequences for the structure of compact stars and for the cooling behavior
of a remnant star after supernova explosion and the subsequent formation of a neutron star.
Thorough understanding on dense nuclear matter is also important
in the physics of relativistic heavy-ion collisions.

In this study, we focus on a specific aspect of dense matter: transition from nuclear matter
 to strange matter.
 This transition is essential to understand relevance of kaon condensation in neutron star.
 This is because the presence of strangeness matter tends to hinder the formation of kaon condensation
 basically due to the Pauli exclusion principle. Here kaon condensation means $K^-$ condensation.
 Since $K^-$ is composed of $\bar u$ and $s$, strange matter will expel the strange quark and so the kaon condensation.
 Moreover, according to previous studies, the critical baryon number density for the nuclear matter to strange matter and
 that for the onset of kaon condensation are not very different from each other, $\sim(2-4)\rho_0$, where $\rho_0$
 is the normal nuclear matter density $\simeq 0.17 ~{\rm fm}^{-3}\sim m_\pi^3/2$.
 In conventional approaches, however,
 when one estimates the transition to strange matter, a bit large uncertainty comes in due to lack of
 robust information on hyperon coupling constants, see~\cite{hyper} for a review.
  In this study, we take a first step towards this direction in holographic QCD.
Recent developments based on AdS/CFT ~\cite{Maldacena:1997re,Gubser:1998bc,Witten:1998qj}
renders a new tool to study dense matter in
the framework of a holographic model of QCD~\cite{TD, EKSS}, see \cite{Erdmenger:2007cm}
 for a review.
There have been many studies on dense nuclear matter \cite{denseMatter} in the holographic QCD.

To study the transition we introduce two flavor D6 branes  which correspond to light (u or d)  and intermediate (strange) quarks
respectively and spherical D4 brane with $N_C$ fundamental strings. The fundamental strings can be attached
 on a light quark D6 brane and/or intermediate mass quark (strange) D6 brane.
 By solving DBI action and applying the force balance
condition, we find stable configuration of D4/D6/D6 system. After finding minimum energy configuration,
we calculate the ratio of up and strange quarks in the system as a function of baryon density.

 \section{Baryon vertex}
In this section we discuss baryon vertex (spherical D4 brane with $N_C$ fundamental strings)
in confining background following \cite{Seo:2008qc}\par
The non-supersymmetric geometry for confining background of D4 in Euclidean signature is given by
\begin{eqnarray}
ds^{2}
&=&\left(\frac{U }{R }\right)^{3/2}\left(\eta_{\mu\nu}dx^\mu dx^\nu + f(U) dx_4^{2} \right)
+\left(\frac{R}{U }\right)^{3/2}\left( \frac{dU^2}{ f(U)} +U^2 d\Omega_4^2\right) \cr
e^\phi&=&g_s\left(\frac{U }{R }\right)^{3/4},\quad F_4 =\frac{2\pi N_c}{\Omega_4}\epsilon_4, \;\; f(U)=
1-\Big(\frac{U_{KK}}{U}\Big)^{3}, \;\; R^3=\pi g_s N_c l_s^3.
\label{adsm}
\end{eqnarray}
This background is related to the black hole solution of D4 brane by the double Wick rotation.
The Kaluza-Klein mass is defined as inverse radius of the $x_4$ direction:
$M_{KK}=\frac{3}{2}\frac{U^{1/2}_{KK}}{R^{3/2}}$.
While $U_{KK}, g_s$, and $R$ are bulk parameters,
$M_{MM}$ and $g^2_{YM}$ are the parameters of the gauge theory.
These are related by
\be\label{consts1}
g_s=\frac{\lambda}{2\pi l_sN_c M_{KK}}, \quad U_{KK}=\frac{2}{9}\lambda M_{KK} l_s^2,
\quad R^3=\frac{\lambda l_s^2}{2M_{KK}} ,\quad \lambda=g_{YM}^{2}N_{c}.
\ee
Introducing a dimensionless coordinate $\xi$;  $\frac{d\xi^2}{\xi^2}=\frac{dU^2}{U^2f(U)}$,
we obtain the background geometry
\be\label{d4bgmetric}
ds^2 = \left(\frac{U }{R }\right)^{3/2}\left(dt^2 +d\vec{x}^2 + f(U) dx_4^{2} \right)
+\left(\frac{R}{U }\right)^{3/2}\left(\frac{U}{\xi}\right)^2\left(d\xi^2 +\xi^2 d\Omega_4^2\right)\, .
\ee
Here $U$ and $\xi$ are related by
$\left(\frac{U}{U_{KK}}\right)^{3/2} = \frac{1}{2}\left(\xi^{3/2}+\frac{1}{\xi^{3/2}}\right).$ \par

A baryon in three-dimensional theory corresponds to the D4 brane wrapping  $S^4$
on which $N_c$ fundamental strings terminate. In this configuration, the background four-form field strength
couples to the world volume gauge field $A_{(1)}$ via Chern-Simons term.\par
The background metric (\ref{d4bgmetric}) can be written as
\be
ds^2 = \left(\frac{U}{R}\right)^{3/2}\left(dt^2 +f dx_4^2 +d\vec{x}^2  \right)
+R^{3/2}\sqrt{U}\left(\frac{d\xi^2}{\xi^2} +d\theta^2 +\sin^2\theta d\Omega_{3}^{2}\right),
\ee
We take $(t,\theta_{\a})$ as a world volume coordinate of D4 brane and turn on the $U(1)$ gauge field on it,
$F_{t\theta}\ne 0$. For simplicity, we assume that the position of D4 brane and the gauge field depends only on $\theta$ i.e.
$\xi=\xi(\theta)$, $A_{t}=A_t(\theta)$, where $\theta$ is the polar angle in spherical coordinates.
The induced metric on the compact D4 brane is
\be\label{d4met}
ds_{D4}^2 = \left(\frac{U}{R}\right)^{3/2}dt^2
+R^{3/2}\sqrt{U}\left[\left(1+\frac{\xi'^2}{\xi^2}\right)d\theta^2 +\sin^2\theta d\Omega_{3}^{2}\right],
\ee
where $\xi' =\partial \xi/\partial \theta$.
The DBI action for single D4 brane with $N_c$ fundamental strings is given by \cite{callan}
\bea\label{bary-d4}
S_{D4} &=& -\mu_4 \int e^{-\phi} \sqrt{{\rm det}(g+2\pi \alpha' F)}+\mu_4 \int  A_{(1)}\wedge G_{(4)} \cr\cr
&=& \t_4 \int dtd\theta \sin^3\theta
\left[-\sqrt{ \o_+^{4/3} (\xi^2 +\xi'^2)-\tilde{F}^2}
+3 \tilde{A}_t \right] \cr
&=& \int dt {\cal L}_{D4},
\eea
where
\bea
\tau_4 &=& \mu_4 \O_3 g_s^{-1} R^{3}\frac{U_{KK}}{2^{2/3}}=\frac{N_c U_{KK}}{2^{8/3}(2\pi l_s^2)} \cr\cr
\tilde{F} &=& \frac{2\pi \a'F_{t\theta}2^{4/6}}{U_{kk}},~~~~~~ \tilde{A}_t=\frac{2^{2/3}}{U_{KK}}\cdot 2\pi\a'A_t
\eea
with $\omega_{\pm} = 1\pm \xi^{-3}$.
The dimensionless displacement is defined as follows;
\bea
\frac{\partial {\cal L}_{D4}}{\partial \tilde{F}}
&=& \frac{\sin^3\theta \tilde{F}}{\sqrt{({\o_-^2}/{\o_+^{2/3}})(\xi^2 +\xi'^2)
-\tilde{F}^2}} \cr\cr
&\equiv& -D(\theta).
\eea
Then the equation of motion for the gauge field is
\be
\partial_{\theta}D(\theta)=-3\sin^3 \theta.
\ee
By integrating above equation, we get
\be\label{displacement}
D(\theta)=2(2\nu -1)+3(\cos\theta -\frac{1}{3}\cos^3\theta),
\ee
where the integration constant $\nu$ determines the number of fundamental sting;
$\nu N_c$ strings are attached at south pole and $(1-\nu)N_c$ strings at north pole.\par
\begin{figure}[!ht]
\begin{center}
{\includegraphics[angle=0, width=0.3\textwidth]{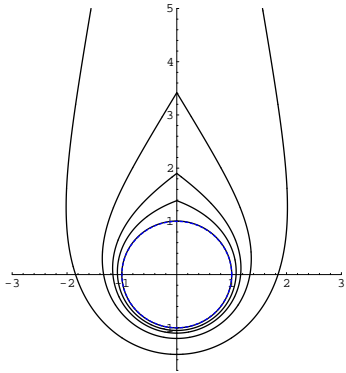}}
\caption{Shape of D4 brane for different $\xi_0$.\label{fig:baryon-d4}
}
\end{center}
\end{figure}

After Legendre transformation, we obtain `Hamiltonian' as
\bea\label{d4h}
{\cal H}_{D4} &=&\tilde{F}\frac{\partial {\cal L}_{D4}}{\partial \tilde{F}}-{\cal L}_{D4}\cr\cr
&=&\t_4 \int d\theta \sqrt{\omega_+^{4/3} (\xi^2 +\xi'^2)}\sqrt{D(\theta)^2+ \sin^6\theta},
\eea
We  solve the equation of motion for ${\cal H}_{D4}$ numerically.  We set $\nu=0$
since we assume that all fundamental strings are attached at north pole $\theta=\pi$.
Then we  impose smooth boundary
condition $\xi'(0)=0$ and $\xi(0)=\xi_0$ at $\theta=0$. Numerical solutions are
parameterized by initial value of $\xi_0$.
The solutions corresponding to different $\xi_0$'s
are drawn in Fig. \ref{fig:baryon-d4}.

If we denote the position of the cusp of D4 brane  by $U_c$, the force at the cusp due to the D4 brane tension
can be obtained by  varying the Hamiltonian of D4 brane with respect to $U_c$ while keeping other variables fixed;
\bea\label{force-d4}
F_{D4}&=& \frac{\partial{{\cal H}}}{\partial U_c} \Bigg|_{\rm fix~other~values} \cr\cr
&=& N_c T_F \left(\frac{1+\xi_c^{-3}}{1-\xi_c^{-3}}\right)
\frac{\xi_c'}{\sqrt{\xi_c^2 +\xi_c'^2}},
\eea
where $ T_F =\frac{2^{2/3}\t_4 }{N_c U_{KK}}$ is tension of fundamental string.
The tension at the cusp of D4 brane is always smaller than the tension of the $N_C$ fundamental strings.
Therefore, if there are no other object, the cusp should be pulled up to infinity  and the final configuration of D4
brane would be `tube-like' shape as in  \cite{callan}.

\section{Holographic transition to strange matter}

\begin{figure}[!ht]
\begin{center}
{\includegraphics[angle=0, width=0.53\textwidth]{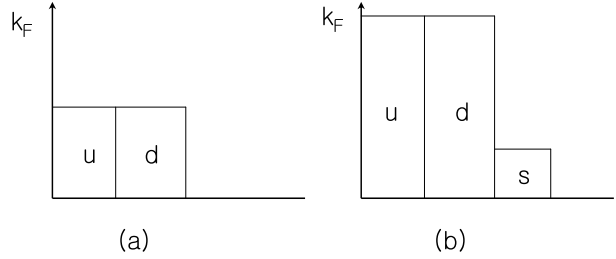}}
\caption{Schematic picture for the nuclear matter (a) to strange matter (b) transition.\label{fig1se}
}
\end{center}
\end{figure}

We begin with an  simplified description of the nuclear matter to strange matter transition.
Figure~\ref{fig1se}  shows a simple view of the transition.
Here we set aside issues like charge neutrality and $\beta$-equilibrium of the dense matter
and consider two light quarks ($u$, $d$ quarks) and one intermediate mass quark ($s$ quark).
The vertical axis of Fig.\ref{fig1se} is roughly the number of quarks in the ground state since $n_q\sim k_F^3$,
where $n_q$ is the quark number density.
In low-density regime, we would have only $u$ and $d$ in our system since the mass of the strange quark, $m_s$, is
roughly a few ten times bigger than that of light quarks, $m_s/m_u\sim 50$, $m_s/m_d\sim 20$,  and so cost too much energy to be piled up in the ground state.
As we increase the number density, the chemical potential of light quarks ($\mu_{u,d}$) become comparable with
the mass of the strange quark, and then system could lower its ground state energy by piling
up some strange quarks if $\mu_{u,d}>m_s$. This argument goes also with baryons.
Instead of transition from $u$, $d$ quark matter to $u$, $d$, and $s$ matter, it could be a transition
from nuclear matter composed of nucleon to strange matter of nucleon and hyperons like $\Lambda$.
As it stands, this is too simple. To be realistic we have to include interaction energy, charge neutrality of the matter,
$\beta$-equilibrium, etc.

Now, we delve into the transition in terms of a holographic QCD.
For simplicity, we will again ignore the charge neutrality and $\beta$-equilibrium of the matter at hand,
relegating them to a future study.
Now we consider the system with two flavors, one light and one intermediate mass quarks.
To introduce two flavors, we put two probe D6 branes in the background.
The the bulk metric (\ref{d4bgmetric}) can be written as
\be
ds^2 = \left(\frac{U }{R }\right)^{3/2}\left(dt^2 +d\vec{x}^2 + f(U) dx_4^{2} \right)
+\left(\frac{R}{U }\right)^{3/2}\left(\frac{U}{\xi}\right)^2\left(d\r^2 +\r^2 d\Omega_2^2+dy^2 +y^2 d\phi^2\right),
\ee
where D6 brane world volume coordinates are $(t,\vec{x},\r,\theta_{\a})$.
The embedding ansatz is that only $y$ depends on $\r$ (we set $\phi=0$). The induced metric on a single D6 brane is
\be
ds^2_{D6} =\left(\frac{U}{R}\right)^{3/2}(dt^2 + d\vec{x}^2) +\left(\frac{R}{U}\right)^{3/2}\left(\frac{U}{\xi}\right)^2
\left[(1+\dot{y}^2)d\r^2 +\r^2 d\O_2^2\right],
\ee
where $\dot{y}=\partial y/\partial\r$.\par
We also introduce $U(1)$ gauge field on D6 brane, which is coupled to the string end point.
The DBI action for the single D6 brane is
\bea
S_{D6}=  \int dt {\cal L}_{D6} &=& -\mu_6 \int e^{-\phi}\sqrt{{\rm det}(g+2\pi \a' F)} \cr\cr
&=& -\t_6 \int dtd\r \r^2  \o_+^{4/3} \sqrt{\o_+^{4/3} (1+\dot{y}^2)-\tilde{F}^2},
\eea
where
\bea\label{consts2}
\t_6 =\frac{1}{4} \mu_6 V_3  \O_2 g_s^{-1} U_{KK}^3 ,\quad
\tilde{F}=\frac{2\cdot2^{2/3}\pi\a'F_{t\r}}{U_{KK}}.
\eea
We define dimensionless quantity $\tilde{Q}$ from the equation of motion for $\tilde{F}$;
\be
\frac{\partial S_{D6}}{\partial \tilde{F}}
=\frac{ \r^2 \o_+^{4/3}\tilde{F}}{\sqrt{\o_+^{4/3} (1+\dot{y}^2)-\tilde{F}^2}}
\equiv \tilde{Q}.
\ee
The Hamiltonian  is connected to the number of point sources (number of fundamental strings) $Q$ by
\be
\tilde{Q}=\frac{U_{KK}Q}{2\cdot2^{2/3}\pi\a'\t_6}.
\ee

 The Hamiltonian can be obtained by the Legendre transformation;
\bea\label{d6h}
{\cal H}_{D6} &=&\tilde{F}\frac{\partial S_{D6}}{\partial \tilde{F}}-S_{D6} \cr\cr
&=& \t_6 \int d\r \sqrt{\o_+^{4/3}\left(\tilde{Q}^2+\r^4 \o_+^{8/3}\right)}\sqrt{1+\dot{y}^2} \cr\cr
&=& \t_6 \int d\r V(\r)\sqrt{1+\dot{y}^2}
\eea
To solve the equation of motion, we impose appropriate initial condition to D6 brane. We are considering
two D6 branes that are connected to a D4 brane with fundamental strings. As discussed in \cite{Seo:2008qc},
the tension of fundamental strings is always larger than that of D-branes. Therefore, two D6 branes are pulled
down and spherical D4 brane pulled up until length of fundamental strings becomes zero. Finally,
the position of the cusp of D6 branes should be located at
 the same position of the cusp of D4 brane $\xi_c$. We consider $Q_1$ fundamental strings attached on one D6 brane
and $Q_2$ strings attached on another D6 brane. We also denote the slope at cusp of each brane as $\dot{y}_c^{(1)}$
and $\dot{y}_c^{(2)}$.
The force at the cusp of D6 branes can be obtained as
\bea\label{force-d6}
F_{D6}&=&\frac{\partial {\cal H}(Q_1)_{D6}}{\partial U_c} \Bigg|_{\partial}
+\frac{\partial {\cal H}(Q_2)_{D6}}{\partial U_c} \Bigg|_{\partial} \cr\cr
&=&\frac{Q_1}{2\pi\a'}\left(\frac{1+\xi_c^{-3}}{1-\xi_c^{-3}}\right)
\frac{\dot{y}_c^{(1)}}{\sqrt{1+\dot{y}_c^{(1)2}}}
+\frac{Q_2}{2\pi\a'}\left(\frac{1+\xi_c^{-3}}{1-\xi_c^{-3}}\right)
\frac{\dot{y}_c^{(2)}}{\sqrt{1+\dot{y}_c^{(2)2}}}\cr\cr
&\equiv& F_{D6}^{(1)}(Q_1)+F_{D6}^{(2)}(Q_2).
\eea
To make the whole system stable, the force at the cusp of D4 brane should be balanced to force of D6 branes;
\be\label{fbc}
\frac{Q}{N_c} F_{D4} =  F_{D6}^{(1)}(Q_1)+F_{D6}^{(2)}(Q_2),
\ee
where $Q=Q_1 +Q_2$.\par
\begin{figure}[!ht]
\begin{center}
\subfigure[]{\includegraphics[angle=0, width=0.45\textwidth]{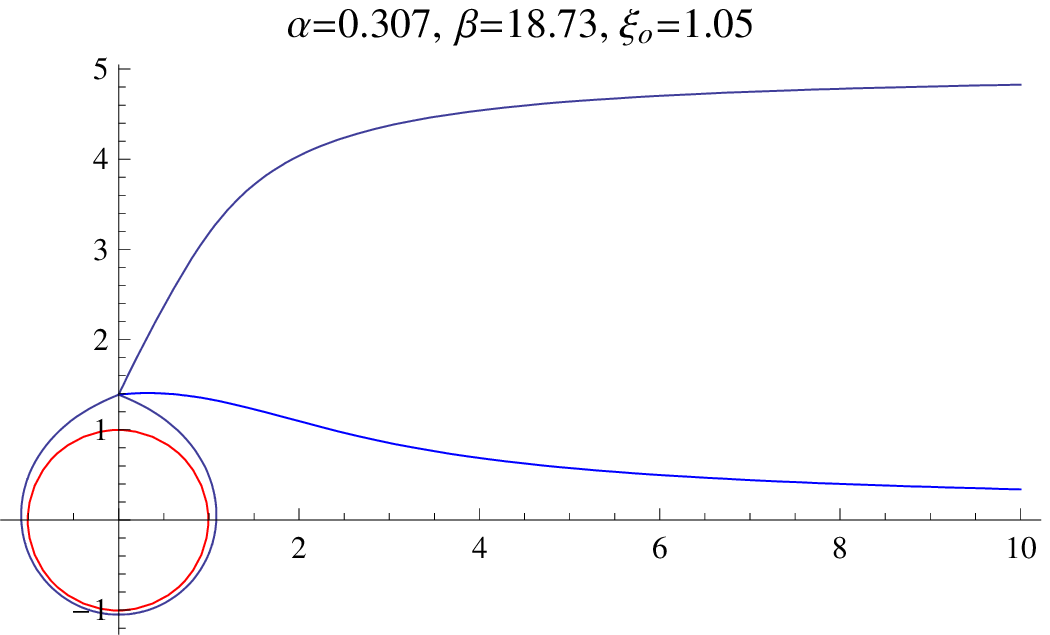}}
\subfigure[]{\includegraphics[angle=0, width=0.45\textwidth]{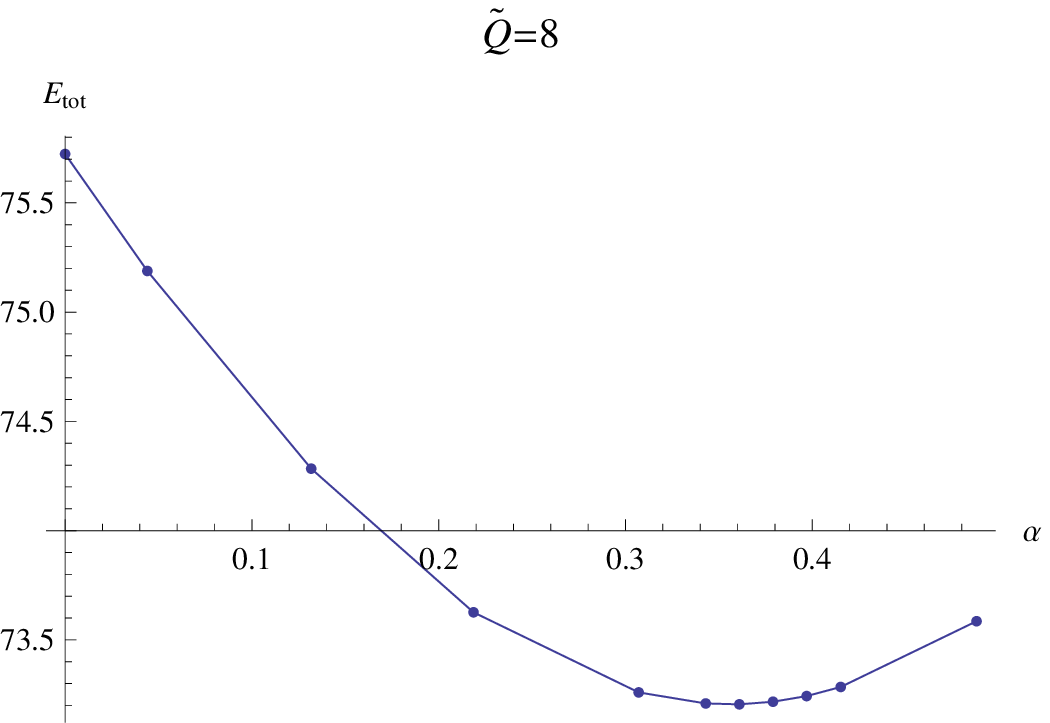}}
\caption{(a) Embedding of D6 branes for $m_{1}=0.1$ and $m_{2}=5$. Red circle
denotes the singularity at $U_{KK}=1$. (b) $\a$ vs. total energy for $\tilde{Q}=8$.\label{fig:HL01}
}
\end{center}
\end{figure}

Rewriting $Q_i$ and $\dot{y}_c^{(i)}$ by using new parameters $\a$ and $\b$,
\bea
Q_1 =(1-\a)Q &,& ~~~ Q_2 = \a Q_2, \cr
\dot{y}_c^{(1)}=\dot{y}_c^{(1)} &,&~~~\dot{y}_c^{(2)}=\b \dot{y}_c^{(1)},
\eea
the force balance condition (\ref{fbc}) becomes
\be\label{fbc2}
\frac{\xi'_c}{\sqrt{\xi_c^{'2} +\xi_c^{2}}} =\frac{(1-\a)\dot{y}_c^{(1)}}{\sqrt{1+\dot{y}_c^{(1)2}}}
+\frac{\a\b\dot{y}_c^{(1)}}{\sqrt{1+\b^2\dot{y}_c^{(1)2}}}.
\ee
We show solutions that satisfy the force balance condition above in Figure.\ref{fig:HL01}(a).\par
Here we fix asymptotic values of D6 branes which are correspond to quark mass to be
$m_{1}=0.1$ and $m_{2}=5$. We call these branes as up and strange quarks brane for convenience.
We note here that  $m_{i}$ is dimensionless and it is related to the real quark mass by the following
relation
\be\label{quarkmass}
M_q = \frac{m_q \lambda M_{KK}}{9 \pi}.
\ee
We will use this relation to discuss the parameter fixing.\par

The behavior of solutions change depending  on the density and $\xi_0$. If we fix $\xi_0$, then from the
equation of motion of D4 brane, the cusp point $\xi_c$ and slope at cusp $\xi_c'$ are automatically determined.
However, from eq. (\ref{fbc2}), for given $\xi_c$ and $\xi_c'$, slopes of two D6 branes
which give $m_{1}=0.1$, $m_{2}=5$ cannot be uniquely determined. In fact, there are infinite set of $\a$ and
$\b$. So we have to choose one embedding by minimum energy condition. The total energy of this system can be
written as follows;
\bea\label{Etot}
E_{tot}&=& \frac{Q}{N_C} {\cal H}_{D4} +{\cal H}_{D6}(Q_1) +{\cal H}_{D6}(Q_2) \cr
&=& \tau_6 \left[ \frac{\tilde{Q}}{4} E_4 +E_6 (\tilde{Q}_1)+E_6 (\tilde{Q}_2)\right],
\eea
where $E_i$ is numerical integration of each `Hamiltonian' in (\ref{d4h}) and  (\ref{d6h}) without overall constant
$\tau_4$ and $\tau_6$. The $\alpha$ dependence of total energy is drawn in Fig. \ref{fig:HL01}(b). In this figure,
we can see  that the total energy of system is minimized at at $\a=0.361$.
In other words, for $\tilde{Q}=8$ case,
 $36.1\%$ of total quarks of ground state (or Fermi sea)
 are occupied by strange (intermediate mass) quarks. For each $\tilde{Q}$, we  find
corresponding $\a$ by imposing the energy minimum condition. \par
\begin{figure}[!ht]
\begin{center}
\subfigure[]{\includegraphics[angle=0, width=0.45\textwidth]{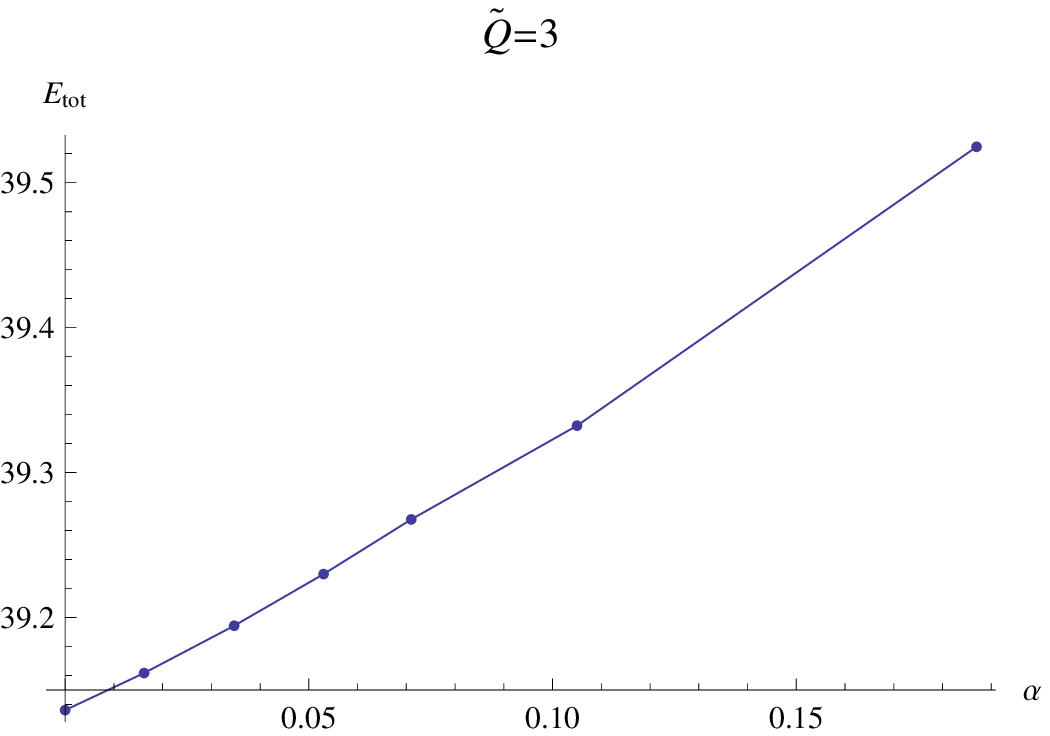}}
\subfigure[]{\includegraphics[angle=0, width=0.5\textwidth]{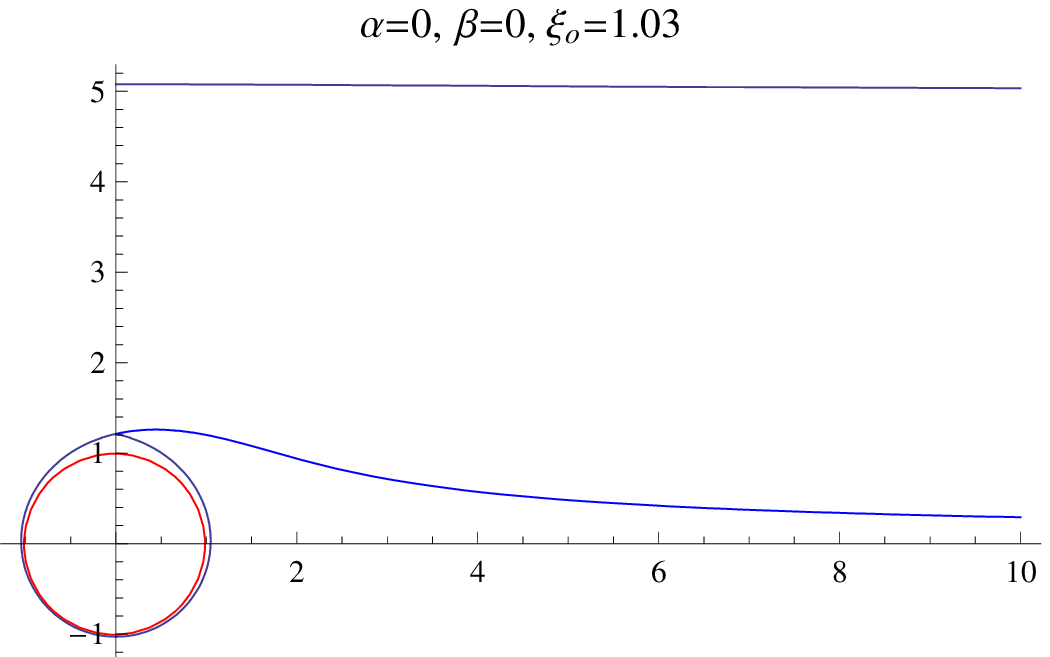}}
\caption{(a) $\a$ vs. total energy for $\tilde{Q}=3$.
(a) Embedding of D6 branes for $\a=0$ with $m_{1}=0.1$ and $m_{2}=5$. Red circle
denotes the singularity at $U_{KK}=1$.\label{fig:HL02}
}
\end{center}
\end{figure}
\begin{figure}[!ht]
\begin{center}
\subfigure[]{\includegraphics[angle=0, width=0.4\textwidth]{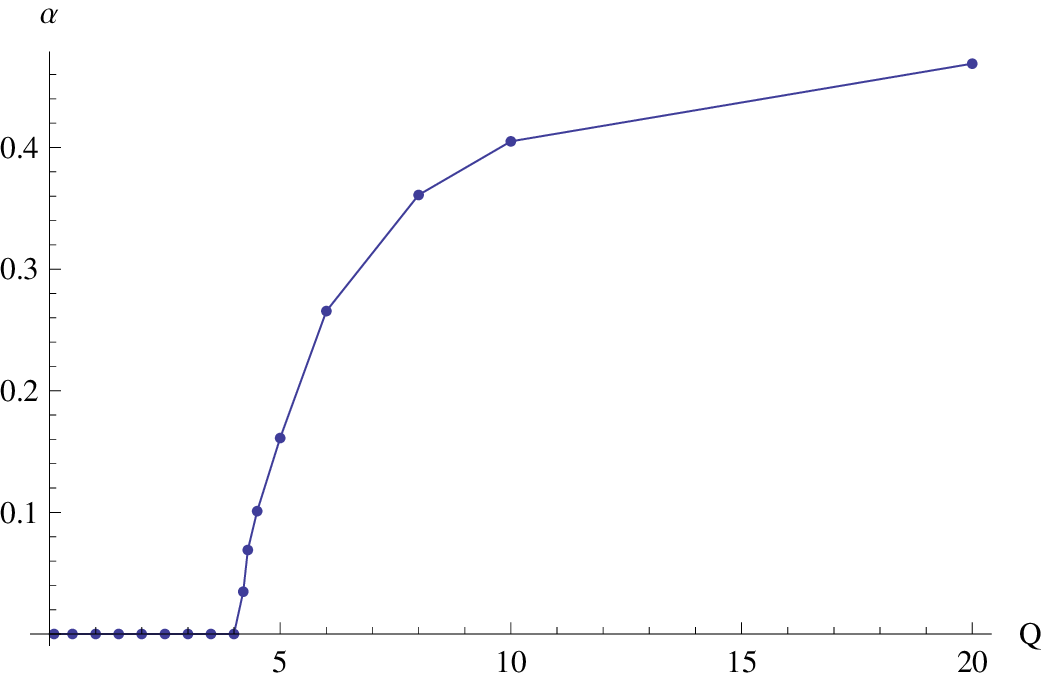}}
\subfigure[]{\includegraphics[angle=0, width=0.4\textwidth]{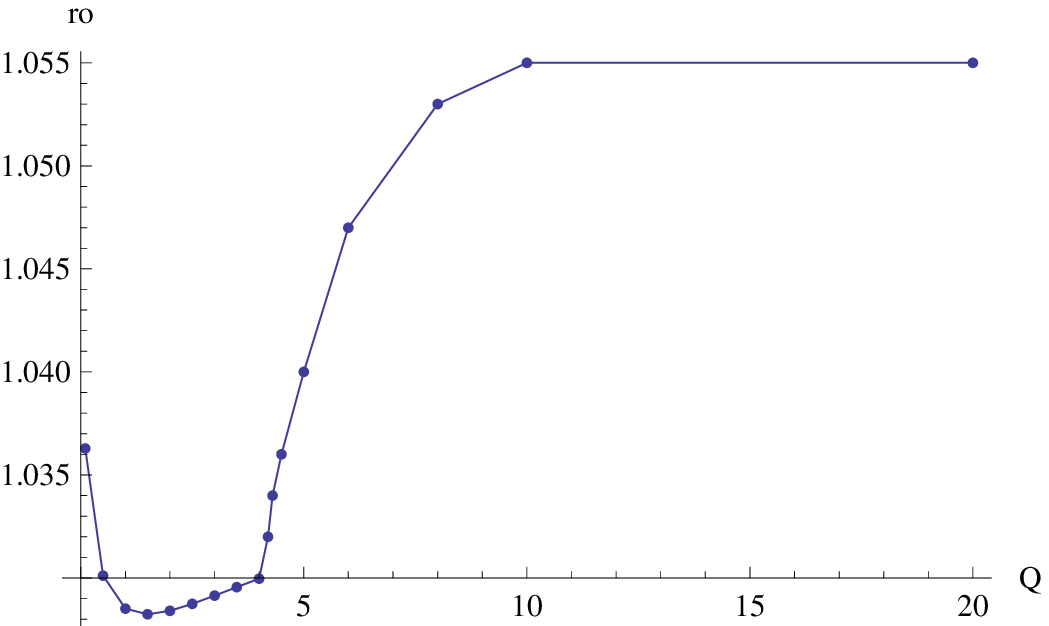}}
\caption{(a) Density dependence of $\a$, the fraction of the strange quarks. (b) Density dependence of $r_0$, a measure of the baryon mass, for $m_{1}=0.1$ and $m_{2}=5$.\label{fig:hla}
}
\end{center}
\end{figure}

For small value of $\tilde{Q}$, $\a$ dependence of total energy have different behavior Fig. \ref{fig:HL01}(b) as shown in Figure.
\ref{fig:HL02}(a). In this figure, total energy   monotonically decreases as $\a$ decreases. Finally, the
minimum energy configuration is at $\a=0$, it means that at small density, strange quark cannot come into the system.
The embedding in this case is drawn in Figure. \ref{fig:HL02}.
As we increase the density, at a certain density, $\a=0$ is not a minimum energy embedding anymore. From $\tilde{Q}\sim 4.2$,
non-zero $\a$ embedding has minimum energy, and the value increases as density increases.
We also check   the large density behavior of embedding: when $\tilde{Q}$ is large, the ratio of strange quark to
up quark seems to go to $0.5$, for example when $\tilde{Q}=500$ the value of $\a$ is around $0.49$.
The final result is drawn in Figure. \ref{fig:hla}(a), which shows sharp transition from nuclear matter ($\alpha=0$) to
strange matter ($\alpha\neq 0$).
It is interesting to note that similar tendency has observed in QCD-rooted model studies, for instance see~\cite{nMsMtQCD}.  \par

Next, we calculate the value of $r_0$  for each embedding which is proportional to the energy of spherical D4 brane.
As we discussed in \cite{Seo:2008qc}, this value can be interpreted as mass of baryon. For small value of $\tilde{Q}$,
the behavior of $r_0$ is the same as \cite{Seo:2008qc} because in this region only one probe brane touch the
baryon vertex - as $\tilde{Q}$ increase, $\r_0$ decreases first and then increases. After $\tilde{Q}\sim 4.2$, two
probe branes are attached to baryon vertex and $r_0$ is increase.
The behavior of $r_0$ as a function of $\tilde{Q}$ is drawn in Figure. \ref{fig:hla}(b).\par
\begin{figure}[!ht]
\begin{center}
\subfigure[]{\includegraphics[angle=0, width=0.4\textwidth]{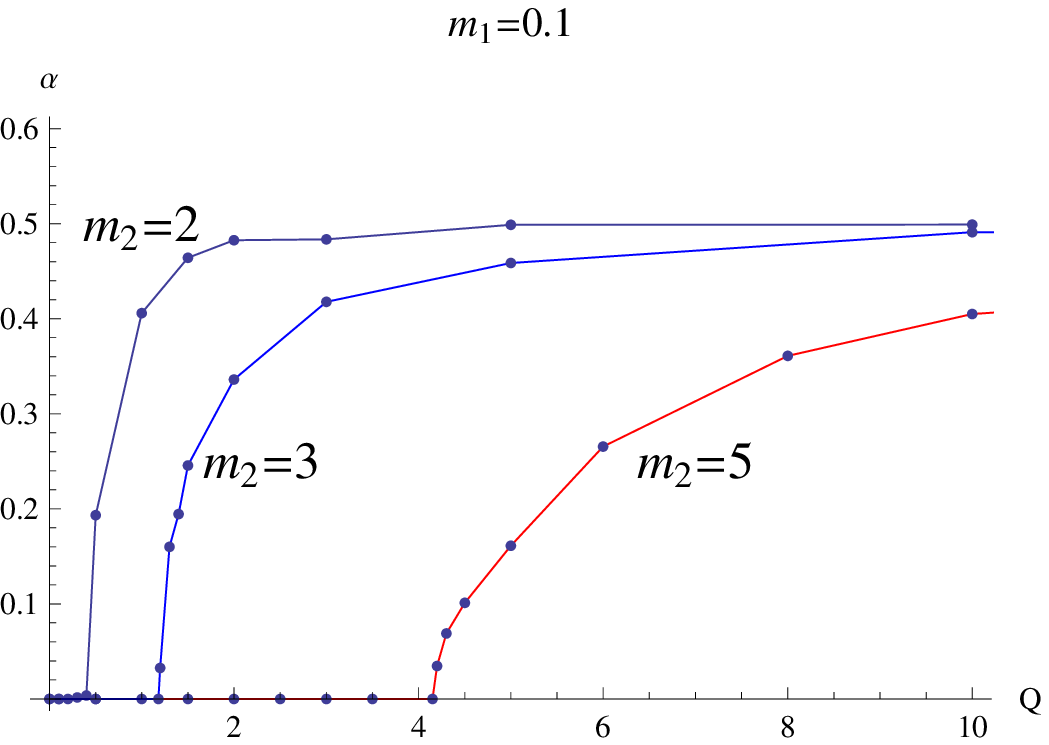}}
\subfigure[]{\includegraphics[angle=0, width=0.4\textwidth]{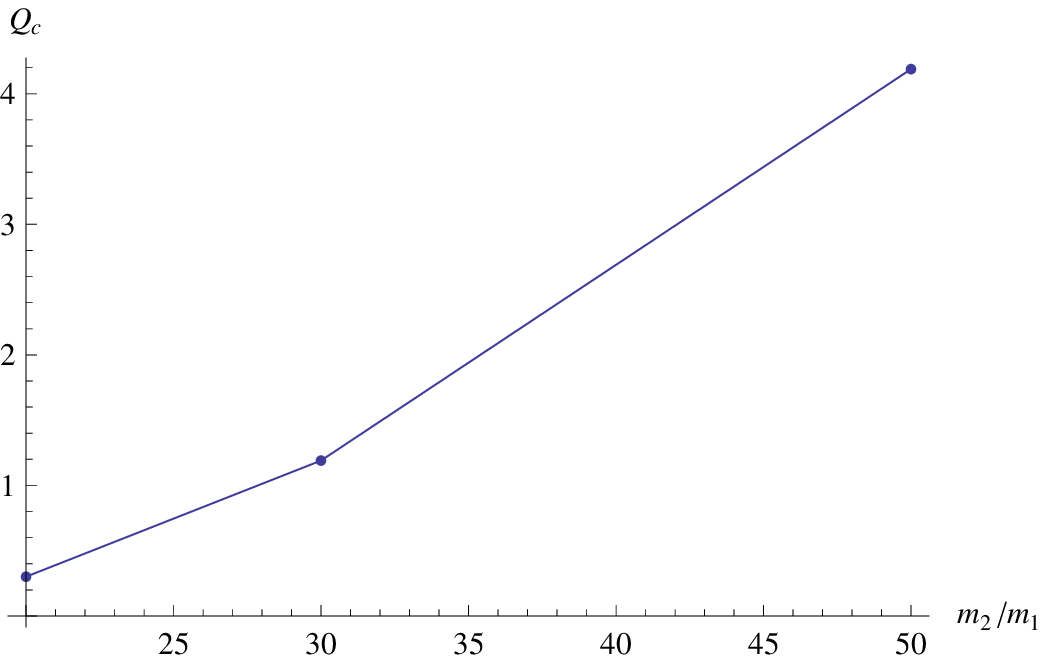}}
\caption{(a) Density dependence of $\alpha$ with different $m_2/m_1$. All lines saturate $\a =0.5$ for
large $Q$. (b) $\rho_c$ vs. $m_2/m_1$.\label{fig:diffm}
}
\end{center}
\end{figure}

Intuitively as $m_2/m_1$ increases, the transition density $\tilde Q_c$ from zero $\a$ to finite $\a$, or from nuclear matter to strange matter,
should increase. We check if our D4/D6/D6 system follows this expectation. For this we used three different values for $m_2$, $m_2=2, 3,5$, with
$m_1=0.1$. The results are summarized in Fig. \ref{fig:diffm} to show that our D4/D6/D6 system respects the intuition.

Though our D4/D6/D6 system is surely far from a realistic dense QCD matter at this stage, we convert the
transition $\tilde Q_c$ obtained in this study into a baryon number density in QCD.
From (\ref{consts1}) and (\ref{consts2}), we can calculate
the baryon number density in terms of $\tilde{Q}$ as follows;
\bea\label{density}
\rho &\equiv& \frac{N_b}{V_3} = \frac{Q}{N_c V_3}\cr
&=& \frac{2\cdot 2^{2/3}}{81 (2\pi)^3}\lambda M_{KK}^3 \tilde{Q}.
\eea
For $m_1=0.1$ and $m_2=5$, we obtained $\tilde Q_c\sim 4.2$.
Empirically the transition baryon number density is $\sim (2-4)\rho_0$, where $\rho_0$ is the normal nuclear density,
from various model studies based on QCD.
To compare our result with the empirical transition density, we {\em choose} $\lambda =1.2$ and $M_{KK}=1.5~{\rm GeV}$.
With this choice we obtain $\rho_c \sim 1.96 \rho_0$ for $\tilde Q_c\sim 4.2$.
From (\ref{quarkmass}), corresponding quark masses are $M_1 \sim 6.4~ {\rm MeV}$ and $M_2 \sim 318 ~{\rm MeV}$,
which are close to the mass of up and strange quarks in QCD.
At the end of the day, however, on completion of our D4/D6/D6 model for realistic dense matter,
we may have to fix the values of $\lambda$ and $M_{KK}$ {\em ab initio} by
considering fluctuations of D6 branes and by doing meson spectroscopy.

\par
\begin{figure}[!ht]
\begin{center}
\subfigure[]{\includegraphics[angle=0, width=0.5\textwidth]{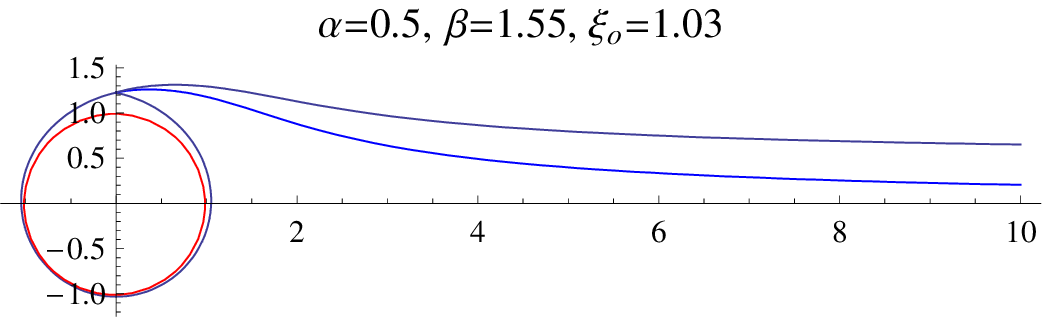}}
\subfigure[]{\includegraphics[angle=0, width=0.4\textwidth]{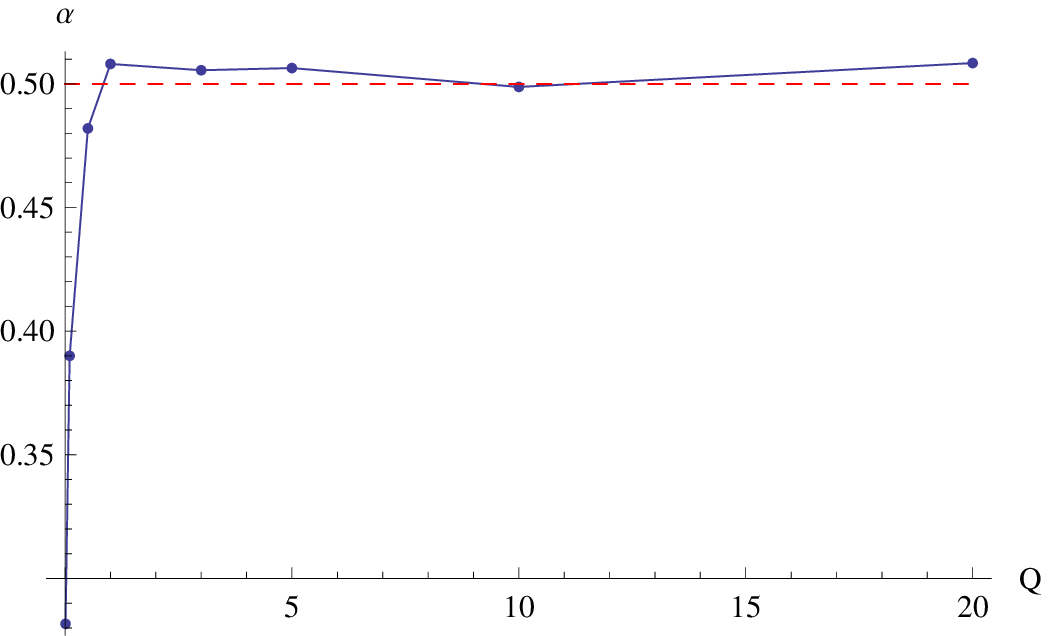}}
\caption{(a) Brane embedding for $m_{1}=0.01$ and $m_{2}=0.5$.
 (b)Density dependence of $\a$.\label{fig:smallqs}}
\end{center}
\end{figure}

So far we consider $m_1=0.1$ and $m_2=2,3,5$.
For fixed ratio between $m_{1}$ and $m_{2}$, however, we can choose several different embeddings.
\begin{itemize}
\item Case 1:  $m_{1}<< U_{KK}=1 <  m_{2}$. This is the case we discussed above.

\item Case 2:   both of $m_q$ are smaller than $U_{KK}$, such  as
$m_{1}=0.01$ and $m_{2}=0.5$.  In this case, brane embedding is drawn in Figure. \ref{fig:smallqs}(a).
The density dependence of $\a$ is drawn in Figure. \ref{fig:smallqs}(b).
In this case, the value of $\a$ seems to saturate to 0.5 for any non-zero density.
That is,  it does not go to zero even in extremely  small density.
We can understand this behavior in geometrically. If both of $m_q$ is smaller than $U_{KK}$, the difference of
geometry detected by each brane is very small. So, the each brane shares   nearly equal number of quarks.

\item Case 3: both of $m_q$ are larger than $U_{KK}$.
Here we  expect similar behavior as in case 2.  Two  quarks will populate evenly.

\end{itemize}
Finally we remark that in our model the value of $\a$ depends on not only the ratio of $m_1$ and $m_2$, but also
each value of $m_1$ and $m_2$.

\section{Summary and Discussion}

In this paper, we constructed a simple holographic QCD model to study nuclear matter to strange matter transition.
The interaction of dense medium and hadrons was taken care of by imposing the force balancing condition for stable D4/D6/D6 configuration.
We considered the oversimplified model where only one intermediate and one light flavor branes are interacting with baryon vertex which is
 homogeneously distributed along $R^3$ space.
We imposed the energy minimization condition and found that  there is a  well defined transition density as a function of current quark mass
and that the transition density increases as the ratio $m_2/m_1$ increases.
We also showed that at very high densities, intermediate (or heavy) and light quarks
populate equally as expected from the Pauli principle. (add a few lines and lower the voice)
So, we may conclude that in our study the effect of Pauli principle is realized as dynamics of D-branes.

We could lessen the oversimplification a  little bit by considering three flavors with $m_u=m_d<m_s$, which could be realized by
considering the light quark favor brane to have weight 2 relative to the intermediate one in the force balancing condition.
That is, if we  change it to
\be\label{fbc2}
\frac{Q}{N_c} F_{D4} = 2 F_{D6}^{(1)}(Q_1)+F_{D6}^{(2)}(Q_2)\, .
\ee
In this case, it is expected that the asymptotic value of $\a$ would be 1/3 instead of 1/2.
To confirm this we take $\tilde Q=500$ and obtain $\a \sim 1/3$.

It will be interesting to consider fluctuations in our D4/D6/D6 system to study meson masses in iso-spin asymmetric matter, which
will be reported elsewhere.

\acknowledgments

Y.K. acknowledges the Max Planck Society(MPG), the Korea Ministry of Education, Science,
Technology(MEST), Gyeongsangbuk-Do and Pohang City for the support of the Independent Junior
Research Group at the Asia Pacific Center for Theoretical Physics(APCTP). The work of SJS was
supported by the WCU project (R33-2008-000-10087-0), KOSEF Grant R01-2007-000-10214-0 and
SRC Program of the KOSEF through the CQUeST with grant number R11-2005-021.
The work of YS is supported by the National Research Foundation of Korea(NRF)
grant funded by the Korea government(MEST) (No. 20090063066).

\end{document}